\documentclass[twocolumn,prl,showpacs,preprintnumbers,superscriptaddress,amsmath,amssymb]{revtex4}     
\usepackage{amsmath,dcolumn,graphicx,epsf,ulem}     
\setkeys{Gin}{width=8.5cm,keepaspectratio}      
\usepackage{bm}% bold math
\begin{document}      
    
\title{The effects of $\rm Co_3O_4$ on the Structure and Unusual Magnetism of $\rm LaCoO_3$}        
         
\author{A.~M.~Durand}         
\affiliation{Department of Physics, University of California, Santa Cruz, CA 95064, USA}   
\author{T.~J.~Hamil}  
\affiliation{Department of Physics, University of California, Santa Cruz, CA 95064, USA}         
\author{D.~P.~Belanger}         
\affiliation{Department of Physics, University of California, Santa Cruz, CA 95064, USA}   
\author{S.~Chi}         
\affiliation{Quantum Condensed Matter Division, Oak Ridge National Laboratory, Oak Ridge, Tennessee 37831, USA}          
\author{F.~Ye}         
\affiliation{Quantum Condensed Matter Division, Oak Ridge National Laboratory, Oak Ridge, Tennessee 37831, USA}       
\author{J.~A.~Fernandez-Baca}         
\affiliation{Quantum Condensed Matter Division, Oak Ridge National Laboratory, Oak Ridge, Tennessee 37831, USA; Department of Physics and Astronomy, University of Tennessee-Knoxville, Knoxville, TN, 37996}     
\author{Y.~Abdollahian}         
\affiliation{Department of Chemistry, University of California, Santa Cruz, CA 95064, USA}      
\author{C.~H.~Booth}          
\affiliation{Chemical Sciences Division, Lawrence Berkeley National Laboratory, Berkeley, CA 94720, USA} %maybe?   
%\author{A.~P.~Ramirez [is he really a contributor?]}         
%affiliation{Department of Physics, University of California, Santa Cruz, CA 95064, USA} %maybe?      
  
\date{\today}          
          
\begin{abstract}          
 
Bulk $\rm La_wCoO_3$ particles with $w=1.1$, $1.0$, $0.9$, $0.8$, and $0.7$ 
were synthesized using starting materials with varying molar ratios 
of $\rm La_2O_3$ and $\rm Co_3O_4$.  The resulting particles are 
characterized as $\rm LaCoO_3$ crystals interfaced with a crystalline $\rm Co_3O_4$ phase. 
X-ray and neutron scattering data show little effect on the average structure and lattice parameters of 
the $\rm LaCoO_3$ phase resulting from the $\rm Co_3O_4$ content, but magnetization data indicate  
that the amount of $\rm Co_3O_4$ strongly affects the ferromagnetic ordering at the interfaces below 
$T_C \approx 89$~K. In addition to ferromagnetic long-range order, $\rm LaCoO_3$ exhibits 
antiferromagnetic behavior with an unusual temperature dependence. 
The magnetization for fields $\rm 20~Oe \leq H \leq 5~kOe$ is fit 
to a combination of a power law $((T-T_C)/T_C)^{\beta}$ behavior representing 
the ferromagnetic long-range order and sigmoid-convoluted Curie-Weiss-like behavior 
representing the antiferromagnetic behavior. 
The critical exponent $\beta=0.63 \pm 0.02$ is consistent with 2D (surface) ordering. 
Increased $\rm Co_3O_4$ correlates well to increased ferromagnetism. 
The weakening of the antiferromagnetism below $T \approx 40$~K is a consequence 
of the lattice reaching a critical rhombahedral distortion as $T$ is decreased 
for core regions far from the $\rm Co_3O_4$ interfaces. 
We introduce a model that describes the ferromagnetic behavior of the interface 
regions and the unusual antiferromagnetism of the core regions. 
  
\end{abstract}           
          
          
\maketitle      %%%Alice editing 9/26/14    
         
\section*{Introduction}   
 
$\rm LaCoO_3$ (LCO), a rhombohedrally distorted perovskite, contains several intriguing 
magnetic features that have been studied for decades without being adequately explained.
Many studies have been done on epitaxial strain in LCO thin films and on surface strain in  
LCO bulk particles. However, systematic studies examining the effects of impurity phase  
interface strain in the bulk particles have been lacking.  In order to explore the effects of 
strain-induced ferromagnetism in this system further, we systematically varied the content 
of $\rm Co_3O_4$ in LCO bulk particles, which allowed us to form a suitable model 
of the magnetic and structural behavior throughout the particles.   

In LCO, a magnetic susceptibility maximum in large fields near $T=40$~K has been 
attributed to a local thermal spin state transition associated with the $\rm Co^{3+}$ 
octahedrally-coordinated ion~\cite{kesaks96,g58,sg95_b}.  
In the CoO$_6$ octahedra, the Co 3d and the O 2p orbitals hybridize to form lower energy  
$t_{2g}$ and higher energy $e_g$ orbitals. Within this localized spin 
excitation model, there has been a long-standing debate as to  
whether the electrons in these orbitals - presumed to be paired in a $S$ = 0, low-spin (LS)  
configuration - excite into either a $S$ = 1, intermediate-spin (IS), a $S$ = 2 high-spin (HS)  
configuration, or a mixture of the two~\cite{le99,plkhy07,fmmpwtvhvg08, bkcc08, inkm94,crbmrsrfg99}. 
In the LS configuration ($t^{6}_{2g}e^{0}_g$), all of the $\rm Co^{3+}$ electrons are  
paired together in the lowest energy orbitals.  
In the HS configuration ($t^{4}_{2g}e^{2}_{g}$), reduction of the crystal field splitting between the  
$t_{2g}$ and $e_g$ energy levels combined with the intra-atomic Hund interaction results in spins able to  
transition into the higher energy $e_g$ levels. In the IS configuration ($t^{5}_{2g}e^{1}_{g}$),  
the splitting is such that only one electron is in the $e_g$ state; this would lead to a doubly  
degenerate ground state resulting in a Jahn-Teller (JT) distortion of two of the  
Co-O bond lengths in the CoO$_6$ octahedron~\cite{kesaks96}.   
  
More recently, the interpretation of the magnetism in the context of local spin state excitations  
has been called into question. Pair-distribution function (PDF) 
and Extended X-ray Absorption Fine Structure (EXAFS) studies showed no significant variation in the  
Co-O bond length, precluding a direct application of the Korotin model.~\cite{jbsbamz09, sjabbbmpz09}.   
Furthermore, band structure calculations suggest  
that the $e_g$ states extend over a large energy range (10 eV), which argues  
for the importance of the extended nature of the electron states~\cite{mlzmfhb12, lh13}.  
We find that the average magnetic moment of $\rm LaCoO_3$ varies with  
temperature in a way more readily interpreted within an extended electron state picture  
emerging from band structure calculations~\cite{lh13} than one where the transitions are between 
spin states of localized Co ions.  An extended, collective  
state model better describes the unusual temperature dependence 
of antiferromagnetic features 
that correlate well with structural changes near $T=40$~K 
and the weak ferromagnetic (FM) long-range order observed for  
$T<89$~K.  
  
The effects of lattice structure and strains on the ferromagnetism of LCO have been considered 
in previous works.  In a study using dynamical mean-field theory in a local density approximation,  
Krapek \textit{et al} found that a magnetic state is favored by lattice expansion and that  
the latter acts as positive feedback for the appearance of local moments~\cite{knknka12}.  
Seo \textit{et al} argued that strained LCO heterostructures can induce thermal spin state 
transitions~\cite{spd12}. By depositing LCO thin films onto different substrates, 
Fuchs \textit{et al} successfully tuned the ferromagnetic ordering 
of LCO;~\cite{fapssl08} an increase in epitaxial strain between  
the substrate and film was observed to correlate with an increase in the effective magnetic moment  
$\mu_B$/Co.  A previous study suggested that ferromagnetic effects are induced by strain from  
Co-impurities in bulk powders of LCO~\cite{dbbycfb13} and that the Co-O-Co bond angle, $\gamma$,  
is a structural indicator of the strength of ferromagnetic ordering.  
Other groups, however, have singled out changes in the Co-O bond  
length~\cite{knknka12, srkkmsw12}, the amount of rhombohedral distortion~\cite{lh13},
or the unit cell volume~\cite{wzwf12} as direct mechanisms for ferromagnetic ordering.  
  
Recently, a spintronic device was constructed by exploiting the strain-induced ferromagnetic  
long-range order in a thin film of LCO.  The piezoelectric properties of a thin $\rm SrTiO_3$  
film allowed voltage switching of the ferromagnetism of the LCO layer~\cite{hppjdy13}.  
Developing a fundamental understanding of the strain-dependent ferromagnetism of LCO and similar  
perovskite structures is a crucial step in identifying other perovskites suitable for 
such spintronic devices and, in particular, will guide the design of devices with ferromagnetic 
materials that can be switched at more practical temperatures.  More comprehensive models 
must address the collective behaviors associated with the strain-switched ferromagnetism. 

%%%%%%%%%%%%Moving this paragraph to near the top%%%%%%%%%%%%%% 
%Many studies have been done on epitaxial strain in LCO thin films and on surface strain in the   
%LCO bulk particles. However, systematic studies examining the effects of impurity phase  
%interface strain in LCO bulk particles have been lacking.  In order to explore the effects of 
%strain-induced ferromagnetism in this system further, we systematically varied the content 
%of $\rm Co_3O_4$ in LCO bulk particles, which allowed us to form a suitable model 
%of the magnetic and structural behavior throughout the particles.   
  
In order to examine the effects of strain on bulk LCO, we varied the amount of 
$\rm Co_3O_4$ phase in the bulk particles  
by varying the starting materials used in the synthesis.  
The resulting powder samples are $\rm La_wCoO_3$,  
with $w = 0.7$, $0.8$, $0.9$, $1.0$ and $1.1$.  The nonstochiometry is not  
expressed uniformly across the samples.  Instead, as will be discussed,  
in addition to large crystalline regions of LCO,  
well-formed crystals of $\rm Co$ oxide-based impurity phases form.  This  
suggests, to a good approximation, that particles consist of regions of  
$\rm LaCoO_3$ interfacing with crystalline impurity phases,  
predominately that of $\rm Co_3O_4$.  The structure and magnetic  
properties of the particles were characterized using  
x-ray diffraction, neutron diffraction, and magnetometry.  
In this report, we review the sample synthesis and discuss the   
sample stoichiometry.  We then present structural results from x-ray and neutron   
diffraction.  Next, we discuss results from SQUID magnetometry.  Finally, we 
discuss a model of the structural and magnetic behaviors in terms of two types of regions, 
interface regions near interfaces between LCO and impurity phases and at 
LCO surfaces and core regions far from such interfaces.  
 
\section*{Synthesis}  
  
The samples were synthesized using a standard solid state reaction. The  
bulk $\rm Co_3O_4$ and $\rm La_2O_3$ were ground together using a mortar and pestle and  
then heated to 850$^\circ$C - 1050$^\circ$C for approximately 8 hours in air. The grinding  
and heating processes were repeated five times, with a final 24 hour heating at 1100$^\circ$C  
in most cases. X-ray powder diffraction characterizations before and after that  
heating showed no differences for the final 24 hour heating; it was not  
applied to the $w = 0.9$ sample used for x-ray diffraction analysis and  
magnetometry. 
  
\section*{X-Ray and Neutron Diffraction}  
  
\begin{figure}  
\includegraphics[width=3.0in, angle=0]{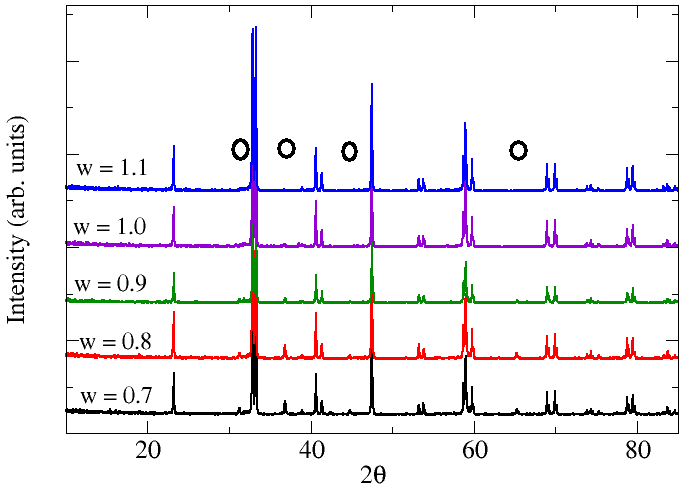}  
\caption  
{X-ray diffraction spectra for all samples at room temperature.  
$\rm Co_3O_4$ structural peak positions are marked by an oval.  
The spectra are vertically offset for clarity.  
}  
\label{fig:XRD_all}  
\end{figure}  
 
X-ray scattering data were taken using a Rigaku 
SmartLab powder diffractometer equipped with a copper x-ray tube 
($\lambda$ (Cu K$_\alpha$) = 1.54056 $\rm \AA$, 
tube energy 44 mA / 40 kV). The samples were analyzed with a 
scan rate of 3.0 $^\circ$/min with a step size of 0.02 $^\circ$.  
Neutron diffraction data were taken at Oak Ridge  
National Laboratory High Flux Isotope Reactor with the US/Japan 
wide-angle neutron diffractometer (WAND), with $\lambda = 1.48 \AA$. Refinements for the  
x-ray data were done using the PDXL software package~\cite{pdxl10} and the ICSD database, 
and those for the neutron data were done using the FullProf program~\cite{r90}. 

\begin{figure}  
\includegraphics[width=3.0in, angle=0]{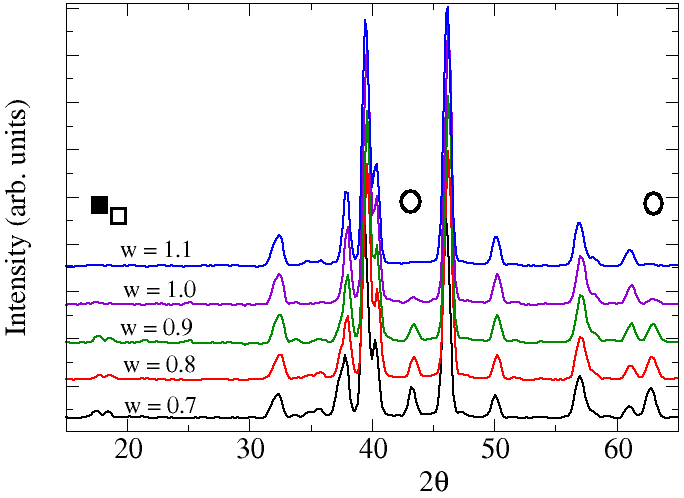}   
\caption  
{Neutron diffraction spectra at wavelength $\lambda = 1.48 \AA$  
for $w = 0.7$, $0.8$, $1.0$, and $1.1$ and $T=30$~K and for $w = 0.9$ at $10$~K.  
The $\rm Co_3O_4$ structural peaks are marked with an oval, the $\rm Co_3O_4$ magnetic   
peak is marked with an empty box, and the CoO magnetic peak is marked with a filled box.  
The spectra are vertically offset for clarity.  
}  
\label{fig:WAND_all}  
\end{figure}  

\begin{figure}  
\includegraphics[width=3.0in, angle=0]{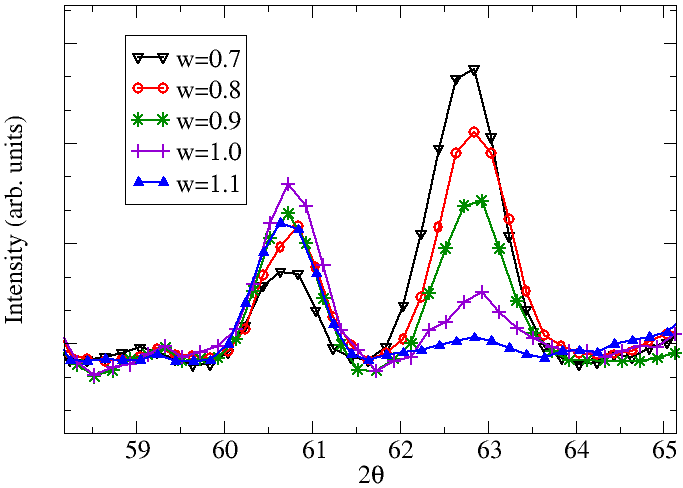}  
\caption  
{$\rm LaCoO_3$ (1 2 5) (left) and $\rm Co_3O_4$ (4 4 0) (right) Bragg peaks   
at room temperature for all samples. The relative peak heights were used, along with  
the concentrations of $\rm Co_3O_4$ in the other samples, to estimate 
the amount of $\rm Co_3O_4$ in the $w = 1.1$ sample.  
}  
\label{fig:Copeak}  
\end{figure}

The x-ray diffraction spectra for all samples at room temperature are shown in Fig.~\ref{fig:XRD_all}. The  
neutron diffraction spectra at $T=30$~K for $w = 0.7$, $0.8$, $1.0$, and $1.1$ and $10$~K for $w = 0.9$   
are shown in Fig.~\ref{fig:WAND_all}.   
Both x-ray and neutron powder diffraction confirm that the samples are predominantly   
$\rm LaCoO_3$, with small amounts of crystalline $\rm Co_3O_4$ phase.  In most cases, 
the $\rm Co_3O_4$ phase was included in refinements.  $\rm Co_3O_4$   
Bragg peak intensities in the $w = 1.1$ sample were too small   
to include in either the x-ray or neutron refinements.  This was also the  
case for the $w = 1.0$ x-ray data.  
Table~\ref{table:percentCo} shows the weight percentage of $\rm Co_3O_4$ in the samples   
as determined by the x-ray and neutron diffraction refinements where possible.  
Figure~\ref{fig:Copeak} shows the relative intensities of the $\rm LaCoO_3$ (1 2 5) %[check that]  
and $\rm Co_3O_4$ (4 4 0) Bragg peaks, which were used to estimate the amount %DEFINITELY 440 
of $\rm Co_3O_4$ in the $w = 1.1$ sample shown in Table~\ref{table:percentCo}.  
  
A magnetic Bragg peak corresponding to a $\rm CoO$ impurity phase is observed 
in neutron scattering scans for all samples below room temperature.  
This is to be expected, because $\rm Co_3O_4$ can convert to $\rm CoO$ at  
temperatures above 900$^\circ$C and the antiferromagnetic ordering temperature of  
the latter is near 290 K~\cite{tkk09}.    
The intensity of the $\rm CoO$ magnetic Bragg peak at 2$\theta$ = 17.2$^\circ$ 
is expected to be significantly larger than the $\rm CoO$ structural ones. 
Because structural peaks corresponding to $\rm CoO$ were not distinguishable in any  
of the scans, the $\rm CoO$ phase were not included in the structural refinements.  
  
\begin{figure}   
   \includegraphics[width=\textwidth, height=2.0in, angle=0]{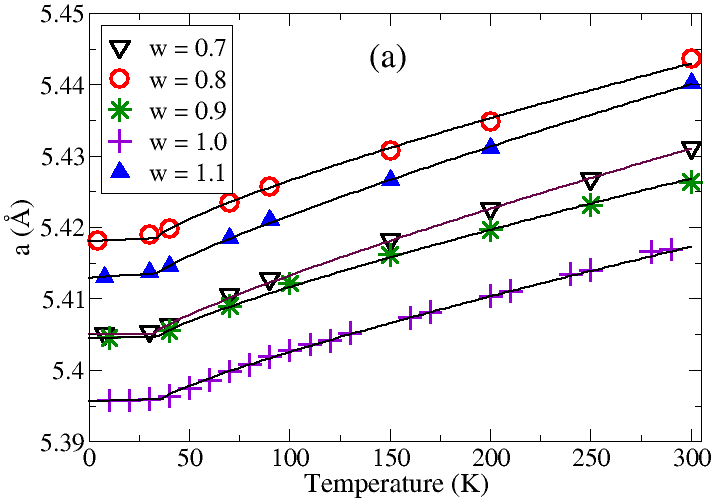}   
   \includegraphics[width=\textwidth, height=2.0in, angle=0]{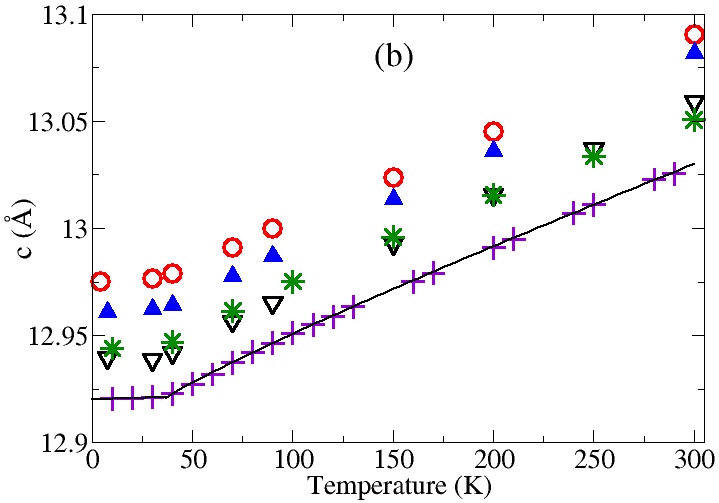}   
   \includegraphics[width=\textwidth, height=2.0in, angle=0]{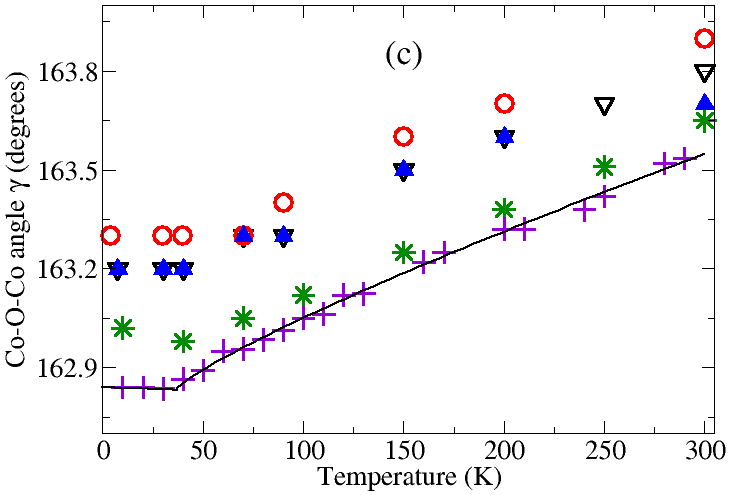}   
   \includegraphics[width=\textwidth, height=2.0in, angle=0]{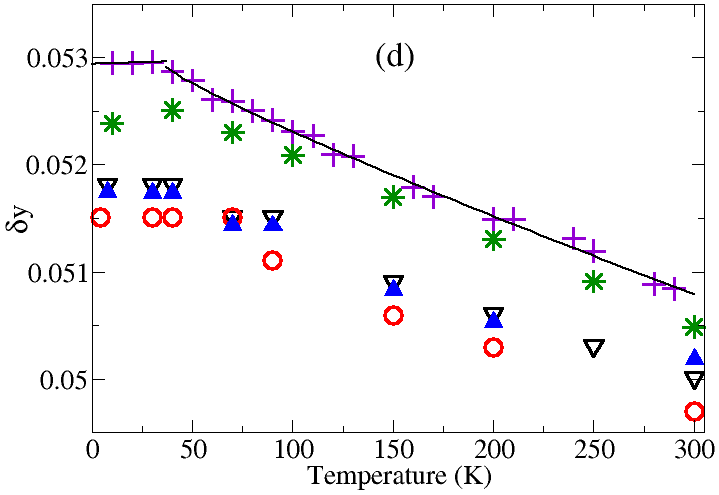}   
 \caption{Lattice parameters $a$ (panel a), $c$ (panel b), the Co-O-Co angle $\gamma$ (panel c),  
and $\delta y$ (panel d) for all samples as determined by neutron diffraction over a  
temperature range of 4 K to 300 K. Fits to power-law behavior for $T \geq T_0$ (Eq.~\ref{eq:power}) and to 
linear behavior for $T \leq T_0$ (Eq.~\ref{eq:linear}) are shown for all samples in (a) and for the 
$w = 1.0$ sample in (b),(c), and (d).~\cite{dbbycfb13}}       
 \label{fig:latticeparams}    
\end{figure}  
  
The formation of $\rm LaCoO_3$ from $\rm La_2O_3$ and $\rm Co_3O_4$ requires  
oxygen supplied by the ambient atmosphere. For example, the most straightforward  
reaction would be  
  
\begin{eqnarray} \label{eq:0}  
  \rm La_2O_3 + \frac{2}{3} Co_3O_4 + \frac{1}{6} O_2 \\  
  \rm \to 2 LaCoO_3 .  
\end{eqnarray}  
  
\noindent{However, the presence of Co impurity phases seen in our $w = 1.0$ sample and  
in previous works on $\rm LaCoO_3$ \cite{mdr67, dbbycfb13} indicates that  
Eq.~\ref{eq:0} does not fully describe reactions that are taking place.  
Likewise, the end products of other $w$ samples do not strictly follow  
straightforward modifications of Eq.~\ref{eq:0}. 
Equation~\ref{eq:0} significantly oversimplifies particle formation 
using the standard synthesis method described above.  

\begin{center}  
\begin{table}   
\caption{Calculated weight percentage of $\rm Co_3O_4$ in each sample at room temperature  
determined using x-ray and neutron scattering data. The value for $w = 1.1$ is estimated  
using the relative intensities of peaks shown in Fig.~\ref{fig:Copeak}.  
}  
\begin{tabular}{ c  |   c   |  c }  
\hline  
\hspace{.5cm} $w$\hspace{.5cm}  &$\rm \% Co_3O_4$ (neutron) & $\rm \% Co_3O_4 (x-ray)$ \\  
\hline  
0.7  &  17.1(0.5)  &  20.3(1.2)  \\  
0.8  &  11.7(0.4)  &  11.8(0.8)  \\  
0.9  &  9.3(0.3)  &  7.4(0.8)  \\  
1.0  &  4.5(0.4)  &  ---  \\  
1.1  &  2.4(0.4)  &  ---  \\  
\hline  
\end{tabular}  
\label{table:percentCo}  
\end{table}  
\end{center}  
  
As examples, two possibilities for the formation of a sample with a 10\% deficiency of $\rm La_2O_3$  
might be envisioned. %probably neither strictly holds. [Why need this part?] 
The first,  
\begin{eqnarray} \label{eq:1}  
\rm \frac{9}{10} La_2O_3 + \frac{2}{3} Co_3O_4 + \frac{19}{60} O_2 \\  
\rm \to 2 La_{0.9}CoO_3 ,  
\end{eqnarray}  
would result in a $\rm LaCoO_3$ perovskite structure with La atoms missing from a significant  
number of sites. These vacant sites would be distributed randomly throughout the lattice, a scenario  
discussed by Androulakis \textit{et al.} \cite{akg01}, and might be accompanied by localized  
structural distortions. The second possibility is the incomplete reaction of $\rm Co_3O_4$  
with $\rm La_2O_3$ in which the $\rm LaCoO_3$ phase forms, but crystalline $\rm Co_3O_4$  
remains in the sample,  
\begin{eqnarray} \label{eq:2}  
\rm \frac{9}{10} La_2O_3 + \frac{2}{3} Co_3O_4 + \frac{3}{20} O_2 \\  
\rm \to \frac{9}{5} LaCoO_3 + \frac{1}{15} Co_3O_4 .   
\end{eqnarray}  
\noindent{The two scenarios require different amounts of oxygen absorbed from the  
ambient atmosphere, with the second process requiring significantly less.  
If absorption of oxygen is a limiting factor, the second would be favored over  
the first.  
  
The observation of minor amounts of crystalline $\rm Co_3O_4$ in our $\rm La_{0.9}CoO_3$ sample   
indicates that the second scenario described in Eq.~\ref{eq:2} might be the better  
description of the synthesis process for our particles.  
$\rm LaCoO_3$ and $\rm La_{1-x}Sr_xCo_3$ powders are often synthesized~\cite{sg95_a, spi03, tth86, wl03, akg01, ewl02, gtpbwl06, gmm96} 
in an air atmosphere and are often described as being single phase as determined by x-ray diffraction. 
When the x-ray diffraction spectra are not shown, it is difficult to assess whether impurity 
phases are present. In a work where the spectra were provided, Ben Amor~\textit{et al.}~\cite{bkcc08}, 
it was reported that varying  
the La:Co ratio resulted in essentially single phase LSCO with a small amount of $\rm CoO$. 
However, the x-ray diffraction spectra exhibit small Bragg peaks corresponding to $\rm Co_3O_4$. 
As we have shown, minute Bragg peaks could represent magnetically 
significant amounts of crystalline $\rm Co_3O_4$. 
We will soon show that even small amounts of this phase can significantly affect the ferromagnetic behavior. 

\begin{table*} 
\caption{Fit parameters for fits to the $\rm La_wCoO_3$ lattice parameters shown 
in Fig.~\ref{fig:latticeparams}. Fits to $a$ were done for all samples, while 
the fits to $c$, $\gamma$, and $\delta y$ are shown for $w = 1.0$ from a previous work.~\cite{dbbycfb13} 
Fits to $a$ were done using the equations 
$a(T) = a_0 + K_a(T - T_0)^{\sigma}$ ($T \geq T_0$) and $a(T) = a(0) + \alpha_aT$ ($T \leq T_0$),
and similar equations were used for $c$, $\gamma$, and $\delta y$. 
The parameters $K_x$ and $\alpha_x$ are in units of $10^{-4}~([x]/\rm K)$ and $10^{-6}~([x]/\rm K)$,
respectively, where $x$ is the relevant lattive parameter and [$x$] are the relevant units. \\
} 
%\centering
\begin{tabular}{c|c|c|c|c|c|c|c}  
  \hline 
  &~~~~~~~$w$~~~~~~~&~~~~~~~$K_a$~~~~~~~&~~~~~~~$T_0$ (K)~~~~~~~&~~~~~~~$\sigma$~~~~~~~&~~~~~~~$a_0 (\rm \AA)$~~~~~~~&~~~~~~~$a(0)~(\rm \AA)$~~~~~~~&~~~~~~~$\alpha_a$~~~~~~~ \\ 
  \hline \hline
  $a$ & 1.1 &  2.30(1)  &  34(3)  &  0.85(2)  & 5.4135(1) & 5.4129(1) & 18.4(1) \\ 
   & 1.0 &  2.08(1)  &  37(2)  &  0.83(2)  & 5.3960(1) & 5.3956(1) & 9.59(1) \\ 
   & 0.9 &  2.15(1)  &  35(3)  &  0.83(2)  & 5.4047(1) & 5.4044(1) & 7.97(1) \\ 
   & 0.8 &  2.80(1)  &  34(3)  &  0.80(2)  & 5.4185(1) & 5.4180(1) & 13.3(1) \\ 
   & 0.7 &  2.37(1)  &  32(3)  &  0.84(2)  & 5.4050(1) & 5.4050(1) & 0.00(1) \\
  \hline \hline
  & $w$ & $K_c$ & $T_0$ (K)~~~& $\sigma$ & $c_0 (\rm \AA)$ & $c(0)~(\rm \AA)$ & $\alpha_c$ \\ 
  \hline
  $c$ & 1.0 & 7.02(1) & 37(2) & 0.91(2) & 12.9206(1) & 12.92(1) & 33.7(1) \\
  \hline \hline
  & $w$ & $K_{\gamma}$ & $T_0$ (K)~~~& $\sigma$ & $\gamma_0 (\rm \AA)$ & $\gamma(0)~(\rm \AA)$ & $\alpha_\gamma$ \\ 
  \hline 
  $\gamma$ & 1.0 & 66.1(1) & 37(2) & 0.84(1) & 162.835(1) & 162.84(1) & 200(1) \\
  \hline \hline
  & $w$ & $K_{\delta y}$ & $T_0$ (K)~~~& $\sigma$ & $\delta y_0 (\rm \AA)$ & $\delta y(0)~(\rm \AA)$ & $\alpha_{\delta y}$ \\ 
  \hline
  $\delta y$ & 1.0 & -0.1558(1) & 37(2) & 0.88(1) & 0.05291(2) & 0.05294(2) & 0.629(1) \\
  \hline \hline
\label{table:lattice_fit} 
\end{tabular} 
\end{table*} 
 
Our $\rm LaCoO_3$ sample contains a small ($\approx 4.5 \%$)  
amount of $\rm Co_3O_4$ phase, which is consistent with insufficient  
oxygen absorption levels to allow complete conversion of the starting materials.  
An example of a reaction utilizing too little oxygen is  
  
\begin{eqnarray} \label{eq:3}  
\rm La_2O_3 + \frac{2}{3} Co_3O_4 + \rm \frac{3}{20} O_2 \\   
\rm \to \frac{9}{5} LaCoO_3 + \frac{1}{15} Co_3O_4 + \frac{1}{10} La_2O_3 .  
\end{eqnarray}  
  
\noindent{This process would result in a sample with 3.4$\%$ $\rm Co_3O_4$ by weight. It must   
also contain $\rm La_2O_3$. However, no $\rm La_2O_3$ Bragg peaks were observed using  
either x-ray or neutron diffraction. It is possible that very small ($\leq 10$ nm)  
nanoparticles of $\rm La_2O_3$ form, which would result in broad, low intensity peaks  
difficult to see in the diffraction spectra. The $\rm La_2O_3$ might also be amorphous.  
  
We examined the effects of an excess $\rm La_2O_3$ phase in the starting   
materials by synthesizing a sample with $w= 1.1$, \textit{i.e.} 10\% excess 
La. One possible reaction could be  
  
\begin{eqnarray} \label{eq:4}  
	\rm \frac{11}{10} La_2O_3 + \frac{2}{3} Co_3O_4 + \frac{1}{6} O_2\\  
	\rm \to 2 LaCoO_3 + \frac{1}{10}La_2O_3 .  
\end{eqnarray}  
  
\noindent{However, although small amounts of $\rm Co_3O_4$ and $\rm CoO$ phases are indicated by the  
Bragg peaks observed in the $w = 1.1$ powder, as in the other samples  
no crystalline $\rm La_2O_3$ phase is observed.  This indicates that the actual growth process is  
likely not homogeneous, as Eq.~\ref{eq:4} would imply.  Inhomogeneous growth processes might produce  
local regions with conditions favoring the formation of $\rm Co_3O_4$.  
 
Figure~\ref{fig:latticeparams} shows the rhombohedral lattice parameters $a$ (panel a)  
and $c$ (panel b) as determined   
from neutron diffraction refinements, as well as the Co-O-Co angle $\gamma$ (panel c) 
and $\delta y$ (panel d). 
For the $w=1.0$ powder, neutron scattering data were obtained for more temperatures  
and with longer counting times.  As a result, there are improved statistics for  
data obtained with this sample relative to the other powders.  
The $\rm LaCoO_3$ structure is a rhombohedrally distorted perovskite (R$\overline{3}$c structure),  
where the Co$^{3+}$ cations are octahedrally coordinated with the O$^{2-}$ anions. In a cubic system,  
the Co-O-Co angle $\gamma$ is 180$^{\circ}$.  
However, in $\rm LaCoO_3$, the oxygen octahedra are twisted in order to accommodate the rhombohedral distortion,  
and $\gamma$ is closer to $163^{\circ}$.~\cite{rc02} 
The parameter $\delta y$ is a measure of the oxygen anion's deviation from the straight line connecting   
two neighboring Co cations, thereby quantifying the amount of distortion from  
cubic.~\cite{rkfk99,dbbycfb13,mkfoiyk99} Mathematically,
 
\begin{equation} 
	\delta y = \frac{d}{a}\cos(\gamma/2) \quad , 
\end{equation} 
 
\noindent{where $d$ is the Co-O bond length and $a$ is the rhombohedral lattice parameter.~\cite{lh13} 
In these data, the $a$ and $c$ lattice parameters, $\gamma$, and $\delta y$ all show similar 
temperature dependencies. Since $d$ varies by less than 0.5 \% over the 
entire range $5<T<300$~K, $\gamma$ and 
$\delta y$ are nearly proportional and essentially equivalent when characterizing 
the rhombohedral distortion and associated magnetic behavior.
Both parameters have been used previously~\cite{lh13, dbbycfb13} so 
we include both in our discussions. 
 
Interestingly, the lattice parameters do not vary monotonically with the amount of $\rm Co_3O_4$ phase.  
The $w=1.0$ behavior represents the smallest values in the average  
lattice parameters $a$, $c$, and $\gamma$, and the largest value of $\delta y$ for all $T$.  
Aside from the vertical shift of data in Fig.~\ref{fig:latticeparams},  
$\rm Co_3O_4$ does not greatly affect the shape of the lattice parameters vs $T$.  
 
Fits to the $a$ lattice parameter are shown for all samples in Fig.~\ref{fig:latticeparams}(a). 
The fits to the $w=1.0$ data were reported previously~\cite{dbbycfb13} and are shown here for 
the $c$, $\gamma$, and $\delta y$ lattice parameters as well (Fig.~\ref{fig:latticeparams}(b), (c), and (d)).  
In the previous work, it was shown that power law behavior describes the  
behavior of the parameters above $T_0=37(2)$~K, while a
linear, nearly constant, behavior describes the parameters  
below $T_0$. The lattice parameter $a(T)$ is thus well modeled by
  
\begin{equation}          
a(T) = a(0) + \alpha_a T  \quad \quad (T<T_0)    
\label{eq:linear}          
\end{equation}  
and  
\begin{equation}    
a(T) = a_0 + K_a(T - T_0)^\sigma \quad \quad (T>T_0).    
\label{eq:power}    
\end{equation} 
 
\noindent{The quantity $a_0$ is the value of $a$ at $T_0$, while $\alpha_a$  
and $K_a$ (units of $\rm \AA/K$) are tunable coefficients.  
Similar equations were used to fit the $c$, $\gamma$, and $\delta y$ lattice 
parameters for $w = 1.0$. 
%The units of $K_x$ and $\alpha_x$ are given by 
%$[x]/\rm K$, where $[x]$ are the units of the relevant parameter, $x$.  
Fitted parameters for Fig.~\ref{fig:latticeparams} (a) through (d) are all given 
in Table~\ref{table:lattice_fit}. 
For $w = 1.0$, the fits done using Eq.~\ref{eq:power} were found to be significantly  
better than fits done based on the Gruneisen formulation using the Einstein  
model for specific heat.~\cite{dbbycfb13,aynctshok94,crbmrsrfg99}  Unlike the 
Gruneisen fits, the power law fits are able to account for the  
sharp increase in slope near $T_0$ and the negative curvature for 50 $\leq T \leq 300$ K. 
Similar qualitative behavior can be seen in the $w \neq 1.0$ samples; they are also 
well fit by the power law. Although Radaelli \textit{et al} 
fit their lattice parameters in the context of thermal excitations, it is difficult to  
assess those fits in the range we are examining because of the sparse temperature  
sampling.~\cite{rc02} In addition, the fits described are in the context of a Jahn-Teller distortion;
as was previously discussed, a significant such effect has not been found in these materials. 

The power law fits we propose imply a phase transition at $T_0$ that is expressed 
in the lattice parameter $a$ for all $w$. 
As can be seen for the $w=1.0$ sample, the same $T_0$ has been used successfully for 
all of the lattice parameters.~\cite{dbbycfb13}
The exponent $\sigma$ gave the best fit for the range $0.80 \leq \sigma \leq 0.85$ across
the samples. During the fitting procedure, a range of $32 \leq T_0 \leq 37$~K was found 
to give the best fit, with all samples $w \neq 1.0$ exhibiting 
a lower value than that for $w = 1.0$.
However, as the temperature sampling was only done every $10$~K, further experiments 
would be required to determine $T_0$ more accurately for $w \neq 1.0$.
Nonetheless, the behavior of the other samples appears similar to that of $w=1.0$, 
but with small vertical offsets. 

%%%%%%%%%%%%%%%%%%%%%%%%% Alice edited everything below at 12:36 PM, 10/13/14 %%%%%%%%%%%%%%
\begin{figure} 
\includegraphics[width=\textwidth, height=4.2in, angle=0]{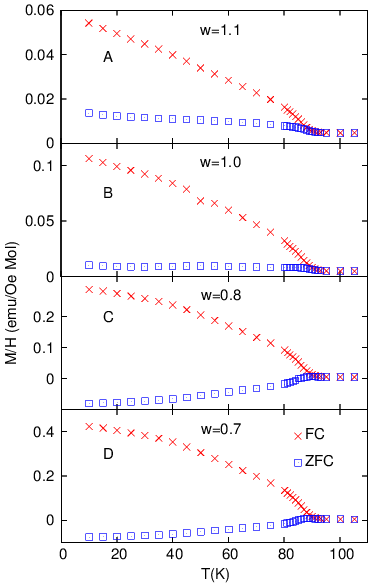} 
\caption{ZFC and FC data for $M/H$ vs $T$ with $H = 20$~Oe for $w=1.1$ (A), $1.0$ (B), 
$0.8$ (C), and $0.7$ (D).} 
\label{fig:FCZFC_20} 
\end{figure} 
 
\begin{figure} 
\includegraphics[width=\textwidth, height=4.2in, angle=0]{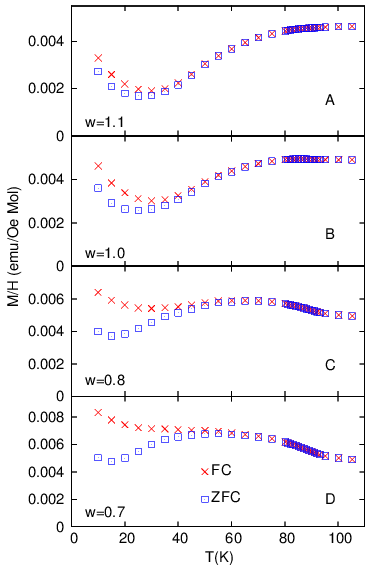} 
\caption{ZFC and FC data for $M/H$ vs $T$ with $H = 5000$~Oe for $w=1.1$ (A), $1.0$ (B), 
$0.8$ (C), and $0.7$ (D).} 
\label{fig:FCZFC_5k} 
\end{figure} 
 
\section*{Magnetometry}  
  
The field-cooled (FC) and zero-field-cooled (ZFC) magnetizations  
were measured in applied fields $20 < H < 5000$~Oe for 10 $\leq T \leq$ 110 K  
using Quantum Design SQUID magnetometers.  
$M/H$ vs $T$ is shown for $w=1.1$, $1.0$, $0.8$, and $0.7$  
in Fig.~\ref{fig:FCZFC_20} and Fig.~\ref{fig:FCZFC_5k} for $H=20$ and $5000$~Oe,  
respectively.  For $H = 20$~Oe, a sudden increase in magnetization indicating  
ferromagnetic ordering is evident upon FC. The critical temperature for this ordering 
is $T_C=88.5\pm 0.5$~K for $w=1.1$ and $T_C=89.5\pm 0.5$~K for $w=1.0$, $0.8$, and $0.7$.  
The ZFC magnetization remains  
small compared to FC below $T_C$ and likely indicates metastable  
domain structure which is frozen in as the sampled is cooled. These ZFC domains are  
not eliminated until $T$ approaches $T_C$.  
A slight difference between FC and ZFC data is observed  
for temperatures $T_C < T < 92$~K, which reflects  
rounding of the transition. This could result from inhomogeneities in the system  
and from the applied field limiting the approach to the  
zero-field ferromagnetic critical point.  
  
For $H=5000$~Oe, the difference between FC and ZFC is smaller  
than for $20$~Oe and increases with the amount of $\rm Co_3O_4$.  
Ferromagnetic long-range ordering is not apparent at this field.  One prominent feature  
of $M/H$ vs $T$ is a ZFC minimum in $M/H$ vs $T$  
near $T=20$~K that becomes more subdued as the $\rm Co_3O_4$ concentration  
increases.  The tendency towards a minimum is weaker for FC, particularly  
for the $w = 0.7$ and 0.8 powders.   
For other values of $H$ between $20$ and $5000$~Oe, the behaviors appear  
intermediate, as expected.  The size of the net moment attributed to ferromagnetic ordering
and the difference between the FC and ZFC magnetizations both increase  
with the $\rm Co_3O_4$ concentrations, indicating that they  
are a direct consequence of this impurity phase.  

\begin{figure} 
\includegraphics[width=\textwidth, height=4.6in, angle=0]{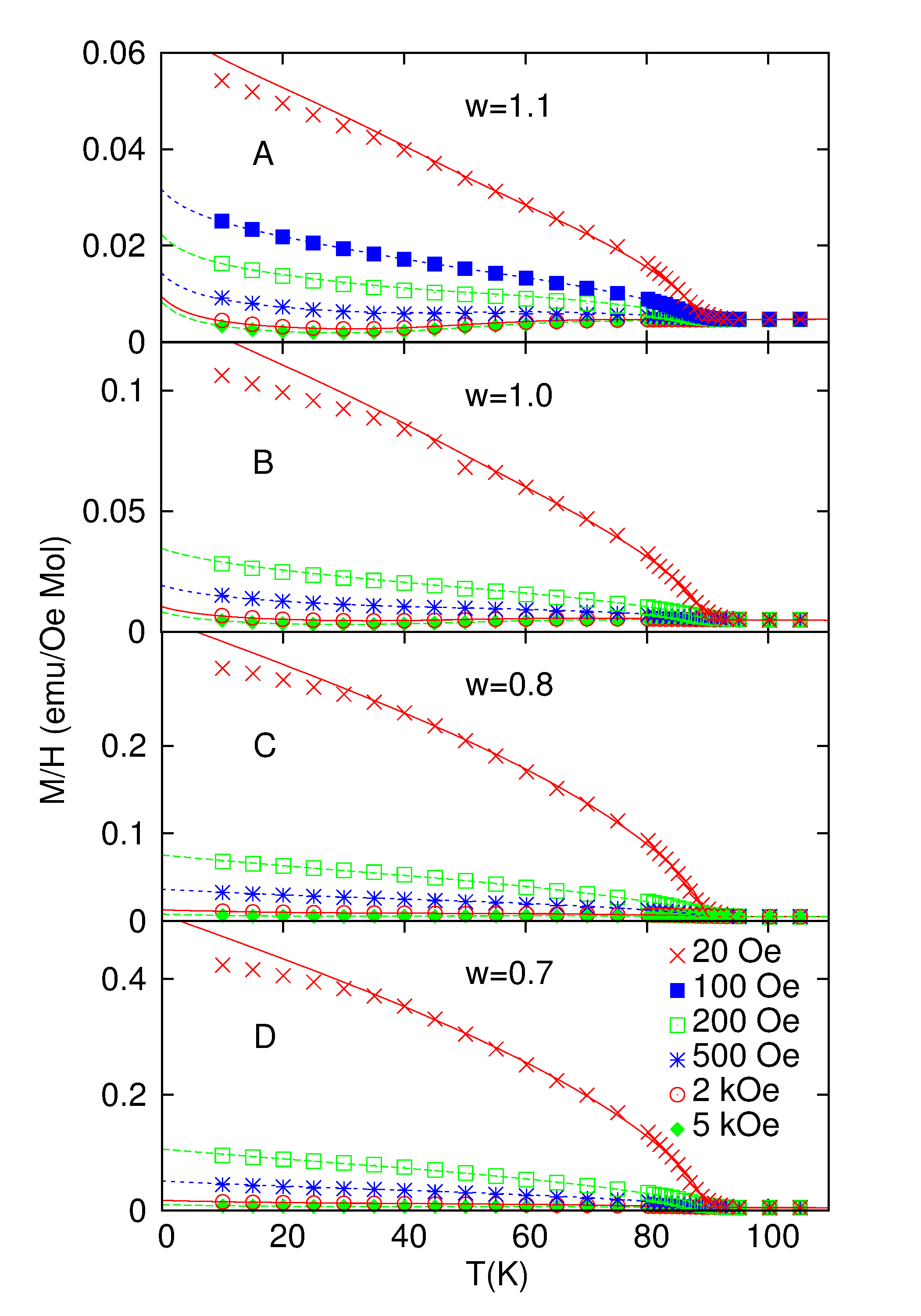} 
\caption{$M/H$ vs $T$ upon FC for $\rm La_wCoO_3$ with $w = 1.1$, $1.9$, $0.8$ and 
$0.7$ in applied fields $20 < H < 5000$~Oe. 
}  
\label{fig:multi} 
\end{figure} 
 
\begin{figure} 
\includegraphics[width=\textwidth, height=4.6in, angle=0]{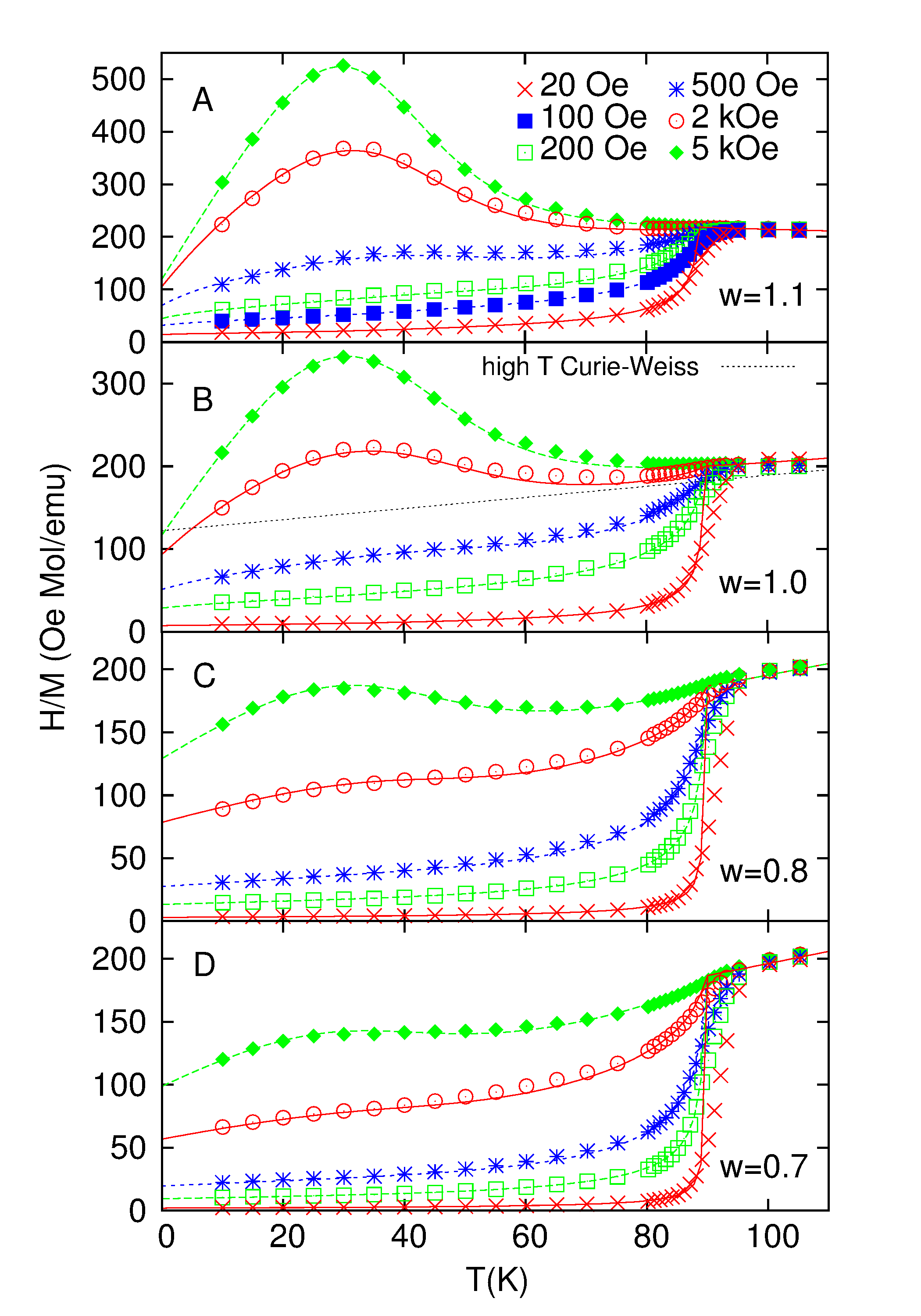} 
\caption{$H/M$ vs $T$ upon FC for $\rm La_wCoO_3$ with $w = 1.1$, $1.9$, $0.8$ and 
$0.7$ in applied fields $20 < H < 5000$~Oe.  The straight line in panel B is the 
Curie-Weiss fit~\cite{dbbycfb13} for $150<T<300$~K. 
}  
\label{fig:multi_inv} 
\end{figure} 
  
Figures~\ref{fig:multi} and \ref{fig:multi_inv}  
show FC data for $M/H$ vs $T$ and $H/M$ vs $T$, respectively, for  
six different applied fields ($H$ = 20, 100, 200, 500, 2000, and 5000 Oe). 
%The high field behavior of all samples is most clearly seen in 
%Figure \ref{fig:multi_inv}.
Well above $T_C$, the behavior appears approximately linear, indicating a paramagnetic Curie-Weiss  
contribution for all samples. For $w=1.0$, the best fit to the  
Curie-Weiss behavior for the range $170<T<300$~K,  
  
\begin{equation}  
H/M = \frac{T+\theta _{CW}}{C}, %Changed from + to -, will change Theta accordingly
\label{eq:CW_fit}  
\end{equation}  
  
\noindent{was done elsewhere~\cite{dbbycfb13} and yielded values  
$C=1.49 \pm 0.02$ emu$\cdot$K/mol and $\theta_{CW} = 182 \pm 4$~K.  
The Curie-Weiss expression gives $H/M = 122 \pm 4$ Oe$\cdot$mol/emu at $T=0$.  
Normally, one would not expect to be able to extrapolate  
to $T=0$, because a transition to antiferromagnetic long-range  
order is expected at a temperature on the order of $\theta_{CW}$, 
and correlated behavior is expected to occur well above that.  
If the behavior followed the straight line indicated 
by Curie-Weiss, the implication would be that correlations remain short-ranged. 
Indeed, the broad peak in $H/M$ near 30 K for all samples at $H$ = 2 and 5 kOe  
indicates significant antiferromagnetic correlations, although the peak 
is much too broad to indicate an antiferromagnetic transition for $T>5K$.  
 
%%% The following paragraph discusses the fits to M/H and H/M 
The data shown in Figs.~\ref{fig:multi} and \ref{fig:multi_inv} indicate  
that there are both ferromagnetic and antiferromagnetic correlations in this  
system. Whereas ferromagnetic long-range order is apparent near 
$T_C$ at low fields, at high fields the system exhibits paramagnetic 
behavior with antiferromagnetic correlations.  A fit of the FC data combining one type of  
paramagnetic behavior and a transition to long-range  
ferromagnetic order for smaller fields proved inadequate. 
In particular, the broad peak for high fields 
in $H/M$ vs $T$, where the contribution from ferromagnetic long-range order is  
very small, is not well characterized by a single paramagnetic-like  
function.  Instead, for $w=1.1$, $1.0$, $0.8$ and $0.7$,  
we were able to fit the FC data to the function  
  
\begin{equation}      
\begin{split}     
\frac{M(T)}{H} =  \bigg(d + \frac{E_a}{T+t_a}\bigg)S(T) + \bigg(\frac{E_b}{T+t_b}\bigg) \\     
+ M_n\bigg(\frac{T_c-T}{T_c}\bigg)^\beta(S(T) + L(1-S(T))),     
\end{split}     
\label{eq:abcd_fit}     
\end{equation}  
 
{\noindent for $T \le T_C$ and 
 
\begin{equation}      
\frac{M(T)}{H} =  \bigg(d + \frac{E_a}{T+t_a}\bigg)S(T) + \bigg(\frac{E_b}{T+t_b}\bigg)     
\label{eq:abcd_high}     
\end{equation} 
 
{\noindent for $T \ge T_C$. The first and third terms of Eq.~\ref{eq:abcd_fit}, along with the 
first term of Eq.~\ref{eq:abcd_high}, are modified by a sigmoid. The sigmoid is characterized 
by an inflection point at a temperature $T_S$ and an inverse width $W$,   
 
\begin{equation}      
S(T) = \frac{1}{1 + \exp(W(T_S-T))} .    %Changing C to T_S because otherwise really confusing with CW parameter 
\label{eq:sigmoid}     
\end{equation}  
  
Equation \ref{eq:abcd_fit} consists of three contributions to the magnetization. The first two terms  
represent paramagnetic behaviors parameterized by Curie-Weiss-like expressions, while the power law behavior  
in the third term represents ferromagnetic long-range ordering below  
the transition temperature $T_C$. The sigmoid function modifies the first term, $E_a/(T+t_a)$, and characterizes  
a reduction of the high temperature paramagnetism over a temperature range parameterized by $W$ and  
centered at a temperature $T_S$. The second term, $E_b/(T+t_b)$, represents the low $T$  
paramagnetic behavior that exists even when the high temperature paramagnetic behavior has  
disappeared. These functions work well over the temperature range $T \leq T_C$.  
However, $E_a$ and $t_a$ differ from the paramagnetic behavior for $T>170$~K because significant 
antiferromagnetic correlations have built up for $T<110$~K.  Similarly, $E_b$ and 
$t_b$ should not be interpreted in the usual way for Curie-Weiss behavior because the 
correlations are not necessarily negligible over this region of the fit. 
The expression should be considered a convenient approximation for fitting the data. 
The quantity $M_n$ is a field dependent coefficient for the ferromagnetic behavior, where $n$ indicates  
the field in Oe. $\beta$ is the power-law exponent. The parameter $L$ 
allows for a slight variation of the power law magnitude as the 
decreasing temperature approaches $T_S$ and improved fits for $w=1.0$ and $1.1$.  
  
The fits of the FC data to Eq.~\ref{eq:abcd_fit} and \ref{eq:abcd_high} 
are shown in Fig.~\ref{fig:multi} and \ref{fig:multi_inv}.  The parameters obtained are  
given in Table~\ref{table:mag_fit}.  Although the fitted curve at each  
field involves a large number of parameters,  
it should be noted that the only parameter that is allowed  
to vary with field is $M_n$.  Hence, the fits are well constrained  
by the data at the several fields used.  The ferromagnetic order is most evident for $H$ = 20 Oe in the  
$M/H$ vs $T$ behavior shown in Fig.~\ref{fig:multi}. For $H/M$ vs T, shown  
in Fig.~\ref{fig:multi_inv}, the high field behavior is most obvious and the  
ferromagnetic order is difficult to discern.   
A good fit for each sample requires that the  
fits for every $H$ be of high quality; this is not necessarily a simple task,
as the behavior of $M/H$ is notably different at high and low fields. 
Data very close to $T_C$ are extremely difficult to fit given the rounding  
of the transition, which can be seen in Fig.~\ref{fig:multi}.  
The paramagnetism is not directly affected by  
the ferromagnetic long-range order, as is evident  
from the additive nature of the two behaviors.  
  
\begin{table}  
\caption{Fit parameters for samples $w$ = 1.1, 1.0, 0.8, and 0.7 using  
Eq.\ \ref{eq:abcd_fit}, \ref{eq:abcd_high} and \ref{eq:sigmoid} with fixed value $\beta=0.63$.  
The subscript $n$ in $M_n$ refers to the applied field in Oe.  Error 
estimates are indicated in the $w=1.0$ column and are similar for the 
other powders except where indicated. 
}  
\centering 
\begin{tabular}{c|c|c|c|c|c|{c}}  
& Units & $w$=1.1 & $w$=1.0 & $w$=0.8 & $w$=0.7 \\  
\hline 
$T_C$ & (K) & 88.5 & 89.5(5) & 89.5 & 89.5 \\  
$W$ & (1/K) & 0.11 & 0.09(1) & 0.09 & 0.09 \\  
$T_S$ & (K) & 47(2) & 49(2) & 45(3) & 40(3) \\  
$t_a$ & (K) & 180 & 180 & 180 & 180 \\  
$t_b$ & (K) &6.5 & 11.5(5) & 40 & 35(5) \\  
$E_a$ & ($\frac{\rm emu \cdot K}{\rm mol \cdot Oe}$) &-0.82& 0.60(2)& 0.86 & 0.85 \\  
$E_b$ & ($\frac{\rm emu \cdot K}{\rm mol \cdot Oe}$) &0.054&0.094(1)& 0.29&0.28(1) \\  
$L$ & -- &1.25&1.2(2)&1&1 \\  
$d$ & ($\frac{\rm emu}{\rm mol \cdot Oe}$) &0.0071&0.0019(2)&0.00&0.0 \\  
$M_{20}$ & ($\frac{\rm emu}{\rm mol \cdot Oe}$) &0.048&0.106(3)&0.337&0.503 \\  
$M_{100}$ &  "  & 0.187&-&-&-\\  
$M_{200}$ &  "  &0.0112&0.022(2)&0.068&0.098\\  
$M_{500}$ &  "  &0.0048&0.0092(3)&0.029&0.043\\  
$M_{2000}$&  "  &0.0009&0.0018(2)&0.0055&0.0095\\  
$M_{5000}$&  "  &0.0000&0.0000(1)&0.0005&0.002\\  
\hline  
\end{tabular}  
\label{table:mag_fit}  
\end{table}  
  
\begin{figure}  
\includegraphics[width=2.5in, angle=0]{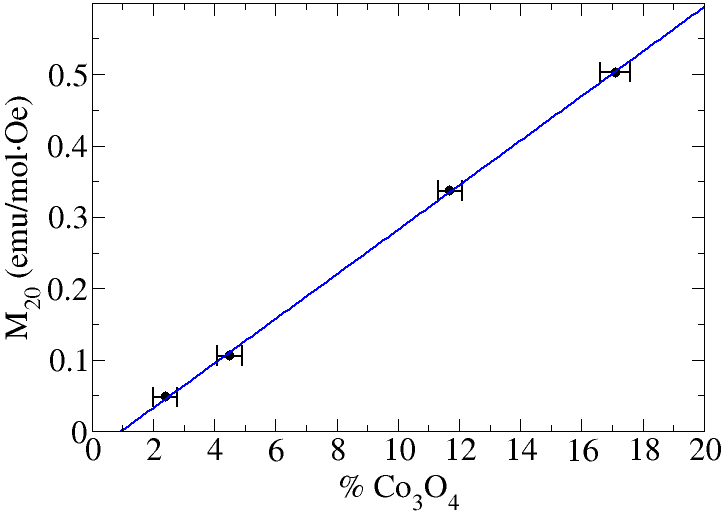}  
\caption  
{The fitted value of $M_{20}$ (see Table~\ref{table:mag_fit}) vs the weight percentage of $\rm Co_3O_4$  
for $w = 0.7$, $0.8$, $1.0$, and $1.1$ (see Table~\ref{table:percentCo}). Weight percentages  
are taken from the neutron scattering only. The data are well fit with a straight line.  
}  
\label{fig:graph_MnvsCo}  
\end{figure}  
  
The magnitude of the power law characterizing the ferromagnetic long-range order  
increases with the concentration of $\rm Co_3O_4$.  Figure~\ref{fig:graph_MnvsCo} shows the  
parameter $M_{20}$, which describes the magnitude of the order parameter  
for $H=20$~Oe, vs the concentration of $\rm Co_3O_4$, determined from  
neutron scattering and listed in Table~\ref{table:percentCo}.  The results are consistent with  
a linear proportionality, as shown in Fig.~\ref{fig:graph_MnvsCo}.   
This behavior indicates that the ferromagnetic  
long-range order is associated with the $\rm Co_3O_4$ particles.  If the  
$\rm Co_3O_4$ particles are of similar size and shape in each of the samples,  
the proportionality is consistent with the amount of $\rm Co_3O_4$/$\rm LaCoO_3$  
interface area. The increase in the ferromagnetic moment at $T_C$ cannot be associated  
with the $\rm Co_3O_4$ particles ordering magnetically, as their antiferromagnetic ordering 
temperature is between $25$ and $40$~K.~\cite{tkk09} 

Further evidence that the ferromagnetism originates within the  
interfaces, and that its critical behavior is dominated by ordering $near$ the  
interfaces, comes from the order parameter critical exponent $\beta$.   
Normally, in three dimensions ($D=3$), the bulk order parameter 
exponent would be close to a value $1/3$, much less than the mean field result $\beta = 1/2$.  
In the case of the ferromagnetism seen  
in the bulk particles, $\beta = 0.63 \pm 0.02$ is clearly greater than  
the mean field exponent.  However, this value is consistent with  
the exponent predicted by Binder and Hohenberg for surface magnetic ordering,~\cite{bh72,bh74}  
which strongly suggests that the ferromagnetism is associated with the surfaces of the 
particles and interfaces with impurity phases. 
Surface ordering is normally assisted by and masked by the  
$D=3$ bulk ordering.  In our case, while the surfaces order ferromagnetically,  
the interior is dominated by the antiferromagnetic interactions and  
never orders.  This allows the ferromagnetic signal to be readily observed, 
but begs the question of how the surface ordering can take place.  
It is possible that in the interface region, antiferromagnetic ordering 
takes place but supports the ferromagnetism via canting of the spins. Another possibility is that 
the antiferromagnetic interactions support the surface ordering even though 
antiferromagnetic long-range ordering never takes place. 
  
\begin{figure}  
\includegraphics[width=3.0in, angle=0]{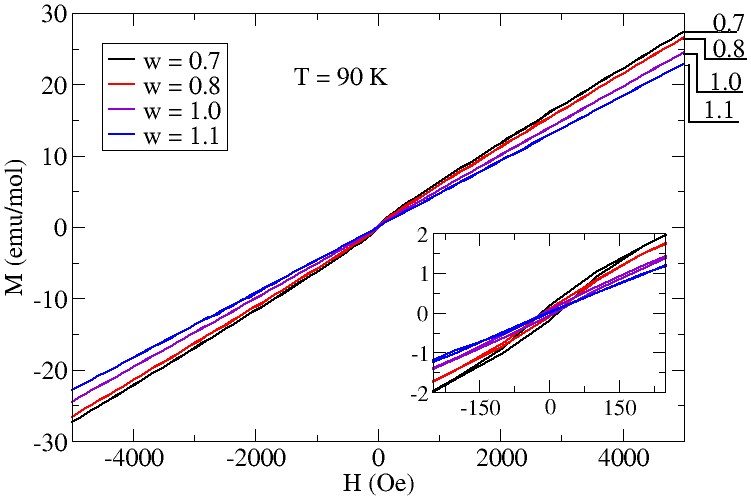}  
\caption  
{$M$ vs $H$ for $w=1.1$, $1.0$, $0.8$ and $0.7$ at $T=90$~K in an applied field $-5 <H< 5$~kOe.  
Inset shows the small amount of hysteresis for near $H = 0$.}  
\label{fig:MvH_90K}  
\end{figure}  
  
\begin{figure}  
\includegraphics[width=3.0in, angle=0]{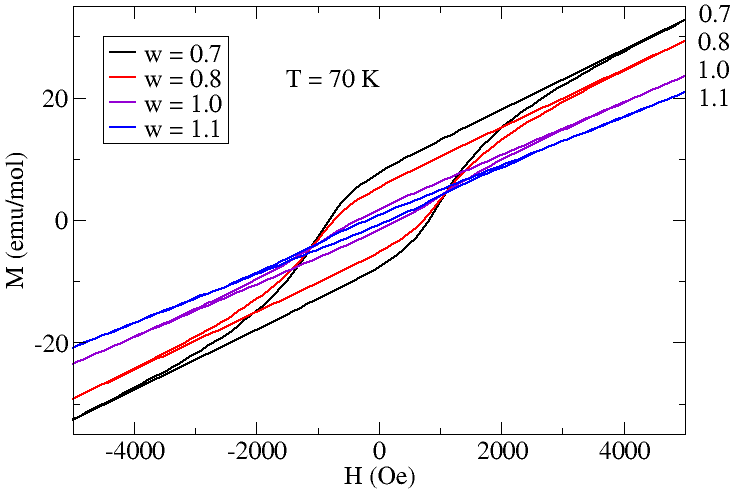}  
\caption  
{$M$ vs $H$ for $w=1.1$, $1.0$, $0.8$ and $0.7$ at $T=70$~K in an applied field $-5 <H< 5$~kOe.  
}  
\label{fig:MvH_70K}  
\end{figure}  
  
\begin{figure}  
\includegraphics[width=3.0in, angle=0]{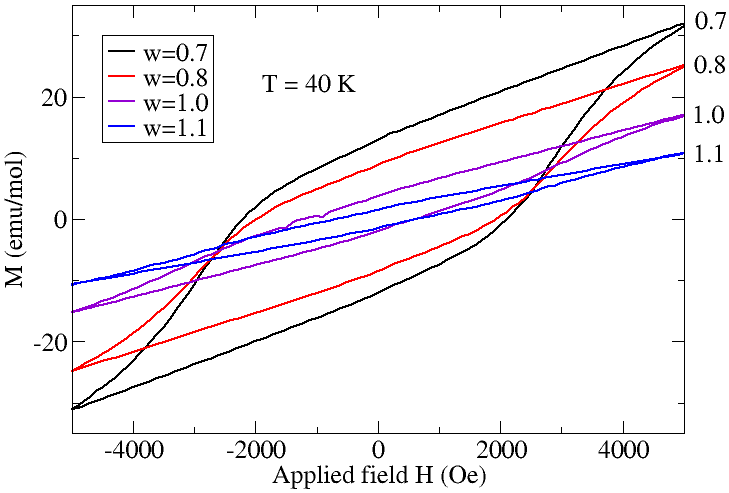}  
\caption  
{$M$ vs $H$ for $w=1.1$, $1.0$, $0.8$ and $0.7$ at $T=40$~K in an applied field $-5 <H< 5$~kOe.  
}  
\label{fig:MvH_40K}  
\end{figure}  
  
\begin{figure}  
\includegraphics[width=3.0in, angle=0]{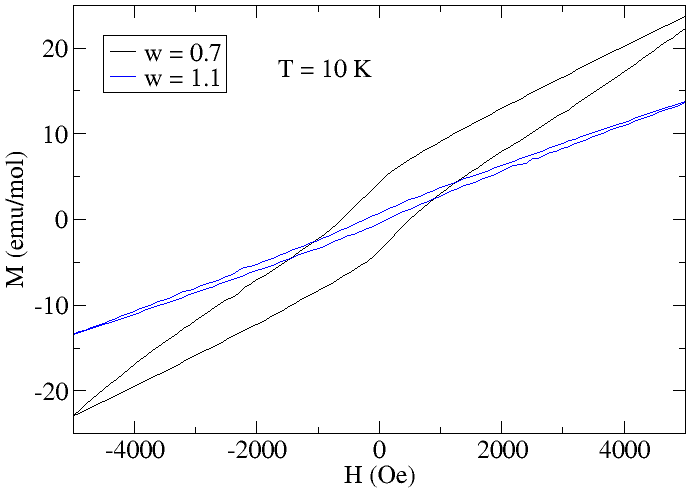}  
\caption  
{$M$ vs $H$ for $w=1.1$ and $0.7$ at $T=10$~K in an applied field $-5 <H< 5$~kOe.  Inset shows the hysteresis for near $H = 0$.  
}  
\label{fig:MvH_10K}  
\end{figure}  
 
\begin{figure} 
\includegraphics[width=3.0in, angle=0]{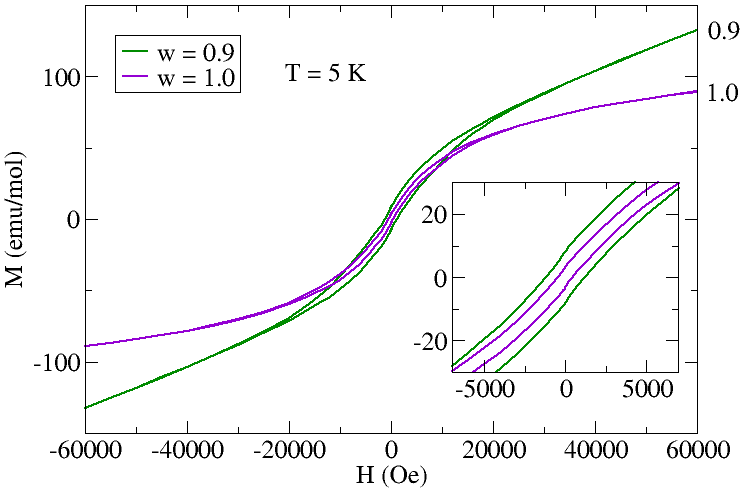} 
\caption 
{$M$ vs $H$ for $w=1.0$ and $0.9$ at $T=5$~K in an applied field $-60 <H< 60$~kOe. 
Inset shows the behavior near $H$ = 0 more clearly. 
} 
\label{fig:MvH_5K} 
\end{figure}

The magnetization curves for $w=0.7$, $0.8$, $1.0$ and $1.1$ as a function of the applied field at  
temperatures $T=90$, $70$, and $40$~K are shown in Figs.~\ref{fig:MvH_90K}, \ref{fig:MvH_70K} and  
\ref{fig:MvH_40K}, respectively (note that different figures have different vertical scales).  
Figure~\ref{fig:MvH_10K} shows similar data for $w=0.7$ and $1.1$ at $T = 10$~K.  
Figure~\ref{fig:MvH_5K} shows data for $w=0.9$ and $1.0$ at $T=5$~K over an applied  
field of $-60<H<60$~kOe.  The samples were ZFC to the measurement temperatures.  The initial  
rise from $H = 0$ to the maximum field is not shown in order to increase the clarity of the plots. 
The values of $M(H=0)$ and $H(M=0)$ are given in Table~\ref{table:hysteresis}.  
 
\begin{table}  
\caption{Values for $M(H=0)$ and $H(M=0)$ for the $M$ vs $H$ plots shown in  
Figs.~\ref{fig:MvH_90K} - \ref{fig:MvH_5K}.
The + and - subscripts indicate whether the value was taken as $H$ was decreased from its  
highest value (+) or increased from its lowest value (-). $M$ is in emu/mol and $H$ is in Oe.  
%**** why isn't M in emu/mol ? ***% 
\\ 
}  
\centering 
\begin{tabular}{c|c|c|c|c|c|{c}}  
$T$ (K) & $w$ & $M_+(H=0)$ & $M_-(H=0)$ & $H_+(M=0)$ & $H_-(M=0)$ \\  
\hline 
5 & 0.9 & 8.11 & -8.14 & 1210 & -1170 \\  
   & 1.0 & 3.25 & -3.16 & 500 & -410 \\  
\hline 
10 & 0.7 & 4.36 & -3.67 & 520 & -620 \\  
   & 1.1 & 0.76 & -0.45 & 190 & -230 \\  
\hline 
40 & 0.7 & 12.9 & -12.1 & 2140 & -2170 \\  
   & 0.8 & 8.83 & -8.51 & 2020 & -1940 \\  
   & 1.0 & 3.75 & -1.95 & 670 & -1130 \\  
   & 1.1 & 1.53 & -1.49 & 740 & -540 \\  
\hline 
70 & 0.7 & 7.74 & -7.65 & 820 & -810 \\  
   & 0.8 & 5.28 & -5.28 & 730 & -730 \\  
   & 1.0 & 1.69 & -1.67 & 330 & -330 \\  
   & 1.1 & 0.77 & -0.78 & 180 & -180 \\  
\hline 
90 & 0.7 & 0.18 & -0.20 & 19 & -17 \\  
   & 0.8 & 0.09 & -0.12 & 13 & -11 \\  
   & 1.0 & 0.07 & -0.08 & 13 & -12 \\  
   & 1.1 & 0.02 & -0.018 & 3.6 & -3.8 \\  
\hline  
\end{tabular}  
\label{table:hysteresis}  
\end{table}  
 
%%%%%%%%%%%%%%%%%%%% Alice stopped editing here, 10/1/14 at 6:02 pm %%%%%%%%%%%%%%%%%%%%% 
  
At $90$~K, all of the samples show a nearly linear response with little hysteresis, as would be   
expected above $T_C$.  The hysteresis is more significant for  
$w = 0.7$ and $0.8$ at $H$ = 0, consistent with the larger ferromagnetic  
moment below $T_C$ in these samples. 
The magnetization at $H=5$~kOe increases by about 4.5 emu/mol as $w$ decreases   
from 1.1 to 0.7, also consistent with a larger ferromagnetic moment  
as the amount of $\rm Co_3O_4$ increases.  
  
At $T=70$~K, well below $T_C$, all samples show significant  
hysteresis (Fig.~\ref{fig:MvH_70K}), while the coercive field $H(M=0)$ and the  
remanent magnetization $M(H=0)$ increase  
with the concentration of $\rm Co_3O_4$, as shown in Table~\ref{table:hysteresis}.  
The samples with $w = 1.0$ and $1.1$ show a smaller net magnetization  
at $H=5$~kOe compared to the values at $T=90$~K.  In contrast,  
for $w = 0.8$ and $0.7$, the magnetization at $H=5$~kOe is larger  
than the values at $T=90$~K. %These behaviors are consistent with increasing  
%antiferromagnetic correlations in the former and ferromagnetic order in the latter [makes sense?]. 
  
For $T = 40$~K, all of the samples exhibit hysteresis (Fig.~\ref{fig:MvH_40K}) loops with a 
much greater enclosed area than for $T=70$ and $90$~K. The hysteresis  
loops extend to the maximum field applied, $|H| = 5$~kOe. The magnetization  
at $5$~kOe is smaller than at higher temperatures for all samples.  
The difference in $M$ between samples at $H=5$~kOe  
is much greater at $T=40$~K than at the higher temperatures, mainly a  
consequence of the decrease in $M/H$ for the samples with less $\rm Co_3O_4$ and the increase for  
those with more $\rm Co_3O_4$.  
 
At $T = 10$~K, hysteresis persists to $H = 5$~kOe for both $w$ = 0.7 and 1.1 (Fig.~\ref{fig:MvH_10K}). 
However, the remanent magnetization in both cases is smaller  
relative to that at the higher temperatures: $M(H = 0)$ of $w$ = 1.1 is half of   
its value at 40 K, and that of $w$ = 0.7 is one third of its value at 40 K.  
Neither sample shows a tendency to saturate for $H < 5$~kOe.  
 
It is clear from Table \ref{table:hysteresis} that the hysteresis loops 
increase as $T$ decreases from $T=90$ to $40$~K, as one might expect as the 
ferromagnetic long-range order increases.  However, it is unusual that 
the hysteresis then strongly decreases as $T$ is reduced to $10$~K. 
This is consistent with the behavior of $M/H$ vs $T$ in Fig.~\ref{fig:multi}, 
where the ferromagnetism drops below the power law for $T<40$~K.   
%This temperature region, $T<T_0$, is also where the antiferromagnetic 
%correlations  weakening -- they weaken at T = 30, not T = 40. 
  
The hysteresis in $M$ vs $H$ at $T=5$~K was measured for $w = 1.0$ and $0.9$ for $-60 \le H \le 60$~kOe. The   
$w = 1.0$ sample appears to be approaching saturation at high fields (Fig.~\ref{fig:MvH_5K}), and  
$M$($H$ = 60 kOe) is 89.3 emu/mol.   
The hysteresis becomes negligible at 40 kOe and the remanent magnetization is only about 0.2 emu/mol   
larger than that in the sample at 40 K. The $w$ = 0.9 sample does not appear to be saturating at 60 kOe,  
and $M$($H$ = 60 kOe) is 133 emu/mol. Although the remanent field is larger in this sample than in the   
w = 1.0 sample (8.1 emu/mol as compared to 3.3 emu/mol), the hysteresis is also nearly gone at 40 kOe.  
These data suggest that there are more ferromagnetically interacting moments available in the w = 0.9 sample 
relative to the 1.0 one, which is consistent with the larger interface region that accompanies 
the higher $\rm Co_3O_4$ content. 
  
%%%%%%%%%%%%%%%% Alice fixed below stuff, 10/13/14 at 12:20 PM %%%%%%%%%%%%%%%%%%%

\section*{Discussion}  
 
As can clearly be seen in Fig.~\ref{fig:FCZFC_20} and \ref{fig:multi}, ferromagnetic long-range order is present  
for all $w$ at low fields. From Fig.~\ref{fig:graph_MnvsCo}, it is clear that the magnitude of the ferromagnetic  
power law contribution increases linearly with the weight percentage of $\rm Co_3O_4$.  
For all samples, this ferromagnetic contribution is well-fit by the exponent $\beta$ = 0.63 
near $T_C$, strongly indicating surface/interface magnetism.~\cite{bh72,bh74}  
Hence, the magnetic data are well modeled with ferromagnetic long-range order 
occuring primarily at the interfaces of LCO and $\rm Co_3O_4$. 

%Antiferromagnetic behavior is superimposed on the ferromagnetism. 
At high fields, where the ferromagnetic long-range order is weak, the antiferromagnetic 
behavior dominates, as can be seen from Fig.~\ref {fig:FCZFC_5k} and \ref{fig:multi_inv} and Table~\ref{table:mag_fit}. 
The gentle increase in $H/M$ just below $T_C$ for $H=5000$~Oe indicates that the antiferromagnetic 
correlations are increasing, as expected from the Curie-Weiss fit for $T \geq 170$~K yielding $\theta_{CW} = 182$~K.
However, there is no indication of a sharp phase transition to long-range antiferromagnetic order. 
Instead, $H/M$ vs $T$ shows a very broad peak near $T=30$~K. The decrease in $H/M$ below this temperature 
appears to indicate a weakening of the antiferromagnetic correlations. 
In general, antiferromagnetic behavior becomes more apparent in the magnetization plots as the $\rm Co_3O_4$ 
amount decreases - this is even more obvious in the single crystal data of Yan \textit{et al},~\cite{yzg04_a} 
where $M/H$ shows a steep dropoff for $T < T_C$. Single crystals likely contain very little $\rm Co_3O_4$
due to the synthesis process.

At $T = 0$, the extrapolated value of $H/M$ is near 
the extrapolated value from the Curie-Weiss fit for $170<T<300$~K. Although possibly coincidental, 
this may suggest that the antiferromagnetic correlations at $T=0$ are insignificant, because 
the high $T$ Curie-Weiss behavior only holds where correlations between spins are insignificant.  
The lack of a transition to antiferromagnetic long-range order and the possibility of negligible 
antiferromagnetic correlations at $T=0$ suggest that the antiferromagnetic interactions are 
frustrated at low $T$.  

The experimental data and the fits performed on the magnetization plots show 
that there exists both ferromagnetic long-range ordering and 
antiferromagnetic correlations in $\rm La_wCoO_3$.  
The fits to the behavior shown in Fig.~\ref{fig:multi_inv} and Table~\ref{table:mag_fit}
indicate that the ferromagnetic and antiferromagnetic behaviors do not interact significantly. 
These findings are consistent with data from other studies~\cite{yzg04_a,fdaeaeksgl09,isnm95,akg01}; 
what remains unclear, however, is the microscopic origin of these behaviors. 
 
Thin film studies~\cite{srkkmsw12,fapssl08,fdaeaeksgl09} have suggested that LCO thin films  
grown on certain substrates exhibit ferromagnetic long-range order, in particular when 
the substrate induces tensile stress at the interface with LCO. 
Sterbinsky \textit{et al} reported that a 20 nm film of LCO deposited on a $\rm SrTiO_3$   
(STO) substrate had a significantly elongated in-plane Co-O bond length and a shortened out-of-plane   
bond length compared to the bulk value.~\cite{srkkmsw12}  They estimated Co-O-Co angles of  
168$^\circ$ and 159$^\circ$ for the in- and out-of- plane bond lengths, respectively. Another  
thin film study by Fuchs \textit{et al} connected structural distortions to the presence of  
ferromagnetic order, estimating that the Co-O-Co angle above which there can be ferromagnetism  
is approximately 160$^\circ$.~\cite{fapssl08,fdaeaeksgl09} All of our data show Co-O-Co angles  
well above this value, with the minimum angle seen at 162.8$^\circ$.  
 
Knizek \textit{et al}, in a more localized spin-state approach,  
discussed how the strong hybridization between the Co $e_g$ states and the O $p$ states might lead to   
the presence of spins in the nominally higher energy magnetic state.~\cite{knj05}  
This hybridization corresponds to a larger   
Co-O-Co angle, a longer Co-O bond length, and a lower amount of rhombohedral distortion (smaller $\delta y$).  
Lee and Harmon, using an extended state GGA calculation, found that for $\delta y$ = 0.052 the magnetic state  
is only 3.2 meV/Co higher in energy than a nonmagnetic state.~\cite{lh13}  
They suggested that even small distortions in the lattice or thermal energy could be enough   
to allow a magnetic state. As $\delta y$ decreases, the minimum in the energy plot 
shifts to favor a magnetic state more strongly.   
Seo \textit{et al} found in their theoretical calculations that tensile stress can induce  
ferromagnetically interacting spins on some Co ions at the film/substrate interface.~\cite{spd12}  
The importance of tensile stress in inducing  
long-range ferromagnetic order is most convincingly demonstrated by the spintronic  
device discussed in Hu \textit{et al}.~\cite{hppjdy13}  By growing a thin film of LCO on a substrate  
of $\rm SrTiO_3$, the piezoelectric property of the substrate can be used to apply  
tensile stress to the LCO film to induce ferromagnetism. 
 
It is clear from these studies that the rhombohedral distortion and 
tensile stress play crucial roles in the magnetic behavior of LCO.  
In our particles, the near proportionality of the amount of $\rm Co_3O_4$ 
and the strength of the ferromagnetic ordering, as well as the 2D 
nature of the critical exponent, strongly suggest that ferromagnetism 
is associated with the interface regions and these regions are a 
source of tensile strain.   
The extensive work performed by Yan \textit{et al} on one sample of LCO with different  
surface areas ranging from a single crystal to nano-sized powder demonstrated that  
the $surface$ of LCO can also be a source of tensile strain and ferromagnetic 
long-range order.~\cite{yzg04_a}  
 
A comprehensive model of the magnetic behavior in LCO must account for the phenomena discussed above, 
namely the surface/interface ferromagnetism and the unusual temperature dependence of 
the antiferromagnetic correlations, including the apparent  
weakening of these correlations below a certain temperature.  
 
Although the ferromagnetic ordering is associated with the interface region, 
it may well propagate into the LCO particles to a significant depth. 
Fuchs \textit{et al} found that highly stressed LCO films exhibited a decreasing ferromagnetic 
order to depths of $100$~nm when deposited on  
$\rm (LaAlO_3)_{0.3}(SrAl_{0.5}Ta_{0.5}O_3)_{0.7}$ (LSAT) substrates.~\cite{fdaeaeksgl09}   
Other experiments indicate ferromagnetism persisting for 20 to 30 nm into  
LCO~\cite{hppjdy13,srkkmsw12}, while experiments on nanoparticle LCO suggest it is in the  
range of $20$ - $100$~nm.\cite{fmmpwtvhvg08_b,httski07}  
Based on these studies, we roughly estimate that ferromagnetism penetrates our 
particles to a depth of 20 to 100 nm. 
 
A significant part of the bulk samples exhibit paramagnetism with net antiferromagnetic 
interactions but with no antiferromagnetic long-range order down to 
our lowest measuring temperature, $T=5$~K.  Furthermore, the correlations appear to 
diminish as $T$ decreases below 30~K. 
Calculations indicate that $\delta y$ correlates with the magnetic state  
and suggest the existence of a critical value near $\delta = 0.052$,~\cite{lh13} very close to the  
critical value $\delta y = 0.053$ found for $w=1.0$.  This critical value 
is achieved just below $T_0$.  Just above it, $\delta y$ and all other lattice parameters 
exhibit a power law behavior in $T-T_0$, which has been said to indicate a phase transition.~\cite{dbbycfb13} 

%%%%%%%%%%%%%%%%% Alice fixed below stuff, 10/13/14 11:51 AM %%%%%%%%%%%%%%%%%%%%
  
To address all of these phenomena we consider a core-interface model consisting of two regions,  
not necessarily sharply delineated or having boundaries that are temperature independent.  
The model addresses all of the observed behaviors listed above.  
The $core$ region describes the particle interiors far from surfaces and interfaces with $\rm Co_3O_4$, while  
the $interface$ region consists of LCO in close proximity to the surfaces and interfaces. The 
surfaces and interfaces impose tensile strain throughout the latter region. 
Particles with fewer surfaces, fewer defects  
that act like surfaces, and fewer interfaces with impurity phases will have relatively  
small amounts of the interface region and more core region.  Note that the model does not address 
the $details$ of the interactions, which we believe are better addressed 
by band structure calculations and experiments such as x-ray absorption and are thus 
beyond the scope of this paper.  
 
We first discuss the interface region. The LCO material near the surfaces and interfaces is strained in 
a manner similar to that in the thin films~\cite{fapssl08,fdaeaeksgl09} and nanoparticles~\cite{httski07} 
in previous studies, all of which exhibit a ferromagnetic transition at $T_C \approx 90$~K. 
The particle size in our bulk materials is essentially the same throughout all of 
the samples, as indicated by the identical synthesis methods and by the lack of variation 
in peak width for the neutron and x-ray scattering data. 
The variation in the amount of ferromagnetism must therefore be due to $\rm Co_3O_4$ in the 
samples, the crystallites of which form interfaces with the LCO during synthesis. 
$\rm Co_3O_4$ is an antiferromagnetic material with $T_N \approx 40$~K~\cite{tkk09},
far removed from the LCO $T_C$. Hence, it is the structural strain and not the 
magnetic ordering of the $\rm Co_3O_4$ that enhances the ferromagnetism in LCO,
in the same manner as a particle surface or a thin film substrate \.
Furthermore, the tensile strain in the interface region 
must extend into the LCO bulk, much like the strain from a substrate extends into the LCO thin film.
Given the significant thermal expansion of LCO (as seen in the lattice parameters, Co-O-Co angle, 
and $\delta y$), the extent of this strain may be affected by the contraction of the lattice as 
the material cools; LCO bulk material in between the core and interface
regions likely experiences a variation in strain with temperature which results in changing magnetic 
correlations.
 
In contrast to the interface region, the core region is dominated by antiferromagnetic interactions,
but never exhibits long-range order. These interactions are strongly affected by the lattice 
contraction with temperature, as can be seen with the rhombohedral distortion parameter.
$\delta y$ increases with a power law behavior as $T$ decreases, until a critical 
value of $\delta y$ between 0.051 and 0.053 is reached at $T_0 \approx 40$~K. Below $T_0$, 
the change is $\delta y$ is relatively small. The transition is 
observed for all values of $w$ studied, although the average $\delta y$ 
is smaller for larger $\rm Co_3O_4$ concentrations. Keeping in mind that 
the values for these lattice parameters are averaged over the entire particle, the variation 
in the low $T$ value of $\delta y$ across samples can be attributed to 
the rhombohedral distortion in the interface region 
remaining smaller than in the core region as a possible consequence of tensile strain. 
As such, only the core material will reach the critical value of $\delta y$, at which 
a magnetic state is not energetically favorable.   
%%% The discussion above discusses the critical angle, but I think the discussion would be simplified  
%%% if we just stuck with delta y (also, with our data in this paper there is no compelling evidence  
%%% for a critical angle anyway). 

%%%%%%%%%%% Alice stopped editing here at 6:12 pm, 10/6 %%%%%%% 

\section*{Conclusion}
 
In conclusion, we have shown that the $\rm LaCoO_3$ phase is robust even in the presence of 
non-stoichiometric starting materials, and that it is unlikely that significant amounts of 
La-deficient unit cells are forming. Instead, the excess $\rm Co_3O_4$ introduced remains 
in the sample in crystalline form and allows for LCO/$\rm Co_3O_4$ interfaces to form.

We introduce an interface-core model as a novel rationalization for the magnetic 
behavior in bulk $\rm La_wCoO_3$. The core LCO material far from the interfaces is paramagnetic 
with antiferromagnetic correlations, and does not order despite a large Curie temperature; this 
is due at least in part to a shift into a non-magnetic state as a result of the system reaching a 
critical value of $\delta y$ as $T$ decreases. 
%As a result of thermal lattice contraction, the rhombohedral distortion is enhanced at low temperatures, 
%effectively preventing long range antiferromagnetic order. 
However, the LCO material near the surfaces and interfaces with $\rm Co_3O_4$ experiences 
tensile strain - this region is thus prevented from reaching the critical value of $\delta y$, and 
in fact exhibits ferromagnetic order with $T_C \approx 90$~K, consistent with the findings of thin film 
studies. Studies such as band structure calculations and x-ray absorption experiments   
are needed to further address the magnetic correlations in the core regions above 
and below $T_o$, as well as the strain mechanism for inducing ferromagnetic long-range order in the 
interface regions

We also present a phenomenological model for the magnetization of $\rm La_wCoO_3$ which is 
consistent with the behavior over a range $\rm 20~Oe \leq H \leq 5~kOe$, and which also fits the 
inverse magnetization. By modifying Curie-Weiss behavior with a sigmoid, and by including 
a ferromagnetic power-law behavior with an exponent indicating surface magnetism, we show that 
the system possesses both antiferromagnetic correlations and ferromagnetic order. Further 
calculations will ideally simplify the mathematical model and further elucidate the origin of 
the sigmoid-type behavior.
 
We thank F. Bridges, A. M. Coogan,        
A. Elvin, B. Harmon, and S. Shastry for        
helpful discussions and$/$or assistance with measurements.         
The work at the High Flux Isotope Reactor at ORNL was supported by 
the DOE BES Office of Scientific User Facilities.       
Work at Lawrence Berkeley National Laboratory was         
supported by the Director, Office of Science (OS),        
Office of Basic Energy Sciences (OBES), of the U.S. Department of         
Energy (DOE) under Contract No. DE-AC02-05CH11231.        
          
\bibliography{magnetism_thesis.bib}          
          
\end{document}